\documentclass[entropy,article,accept,moreauthors,pdftex,12pt,a4paper]{mdpi} 

\pdfoutput=1

\usepackage{graphicx}
\usepackage{url}

\lastpage{5579}
\doinum{10.3390/e15125565}
\pubvolume{15}
\pubyear{2013}
\history{Received: 21 October 2013; in revised form: 12 December 2013 / Accepted: 12 December 2013  / Published: 16 December 2013}

\Title{Global Inequality in Energy Consumption from 1980 to 2010}
\Author{Scott Lawrence, Qin Liu, and Victor M.~Yakovenko *}
\address{Department of Physics, University of Maryland, College Park, MD 20742-4111, USA; \\ E-Mails: srl@umd.edu (S.L.); qinliu.christoph@gmail.com (Q.L.)}
\corres{E-Mail: yakovenk@physics.umd.edu; \linebreak 
Web: \url{http://physics.umd.edu/~yakovenk/econophysics/}}
\abstract{We study the global probability distribution of energy consumption per capita around the world using data from the U.S.\ Energy Information Administration (EIA) for 1980--2010.  We find that the Lorenz curves have moved up during this time period, and the Gini coefficient $G$ has decreased from 0.66 in 1980 to 0.55 in 2010, indicating a decrease in inequality.  The global probability distribution of energy consumption per capita in 2010 is close to the exponential distribution with $G=0.5$.  We attribute this result to the globalization of the world economy, which mixes the world and brings it closer to the state of maximal entropy.  We argue that global energy production is a limited resource that is partitioned among the world population.  The most probable partition is the one that maximizes entropy, thus resulting in the exponential distribution function.  A consequence of the latter is the law of 1/3: the top 1/3 of the world population consumes 2/3 of produced energy.  We also find similar results for the global probability distribution of $\rm CO_2$ emissions per capita.}

\begin{document}

\vspace{-12pt}

\section{Introduction}

\begin{figure}[H]\hfill
\includegraphics[width=\linewidth]{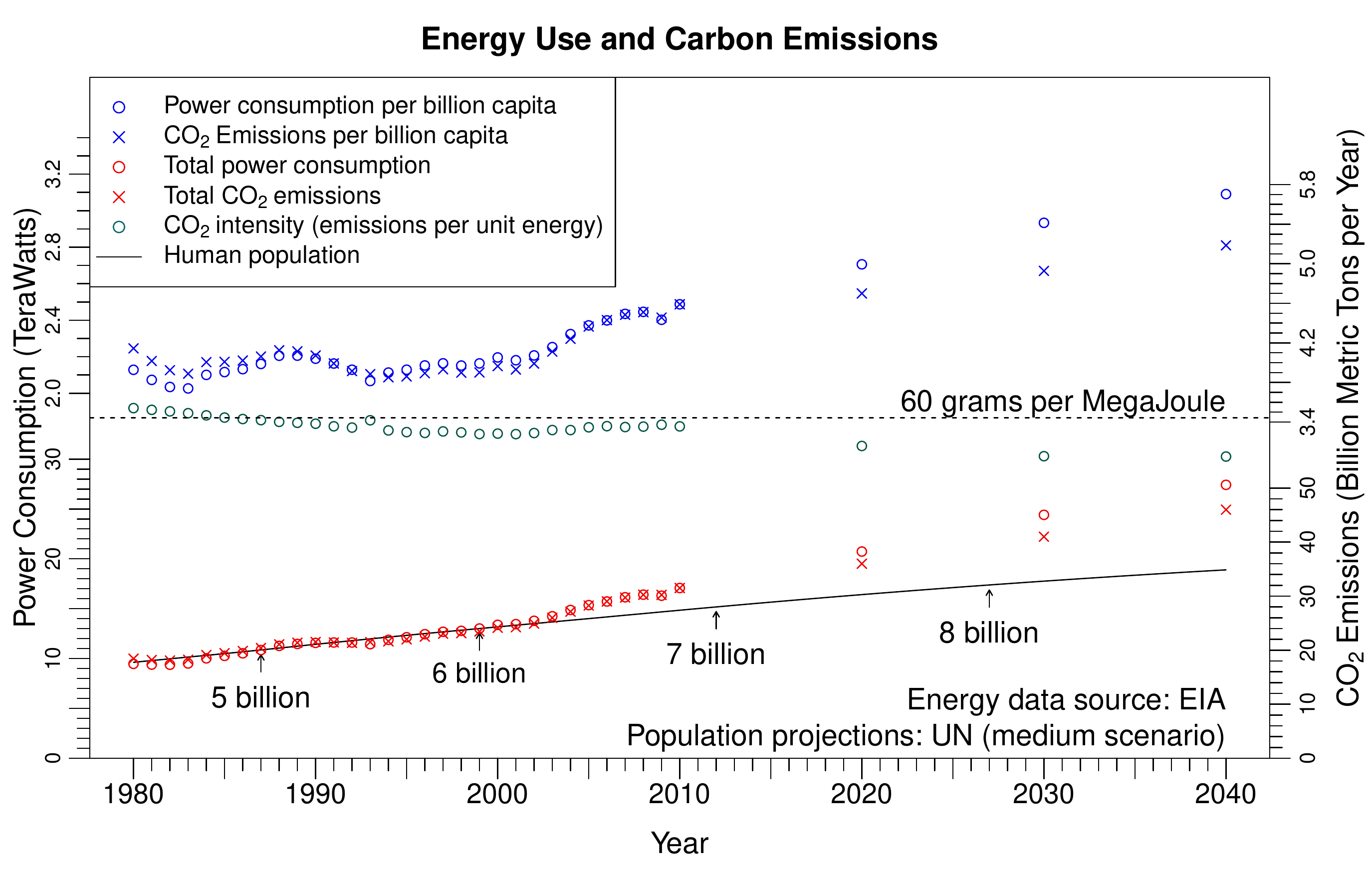}\hfill
\caption{Historical trends (1980--2010) and future projections (2020--2040) for total and per-capita energy consumption (red and blue circles), as well as $\rm CO_2$ emissions (red and blue crosses), along with $\rm CO_2$ intensity (green circles) and human population (black curve).}
\label{Fig:global-trends}
\end{figure}

Energy, climate, and the environment are the biggest problems of our time.  Much of the discussion around these issues focuses on total energy consumption and $\rm CO_2$ emissions by the whole world, as well as the corresponding per capita numbers for the whole world population.  These trends are shown in Fig.~\ref{Fig:global-trends}, where the points for 1980--2010 represent historical data, whereas the points for 2020--2040 are projections by the U.S.\ Energy Information Administration (EIA).  The data source for our paper is Ref.~\cite{EIA}, and spreadsheets for all data plots shown in the paper are available as the supplementary online material.  The upward projections in Fig.~\ref{Fig:global-trends} extrapolate the trend observed from 2000--2010.  However, limiting global warming to 2$^\circ$C requires limiting total cumulative carbon emissions to one trillion tonnes \cite{trillion}, as stated in the 2013 report by the Intergovernmental Panel on Climate Change (IPCC) \cite{IPCC}.  This is a challenging goal, given that the limit is expected to be reached around 2040 based on current trends \cite{NYTimes}.

\begin{figure}[H]
\includegraphics[width=0.50\linewidth]{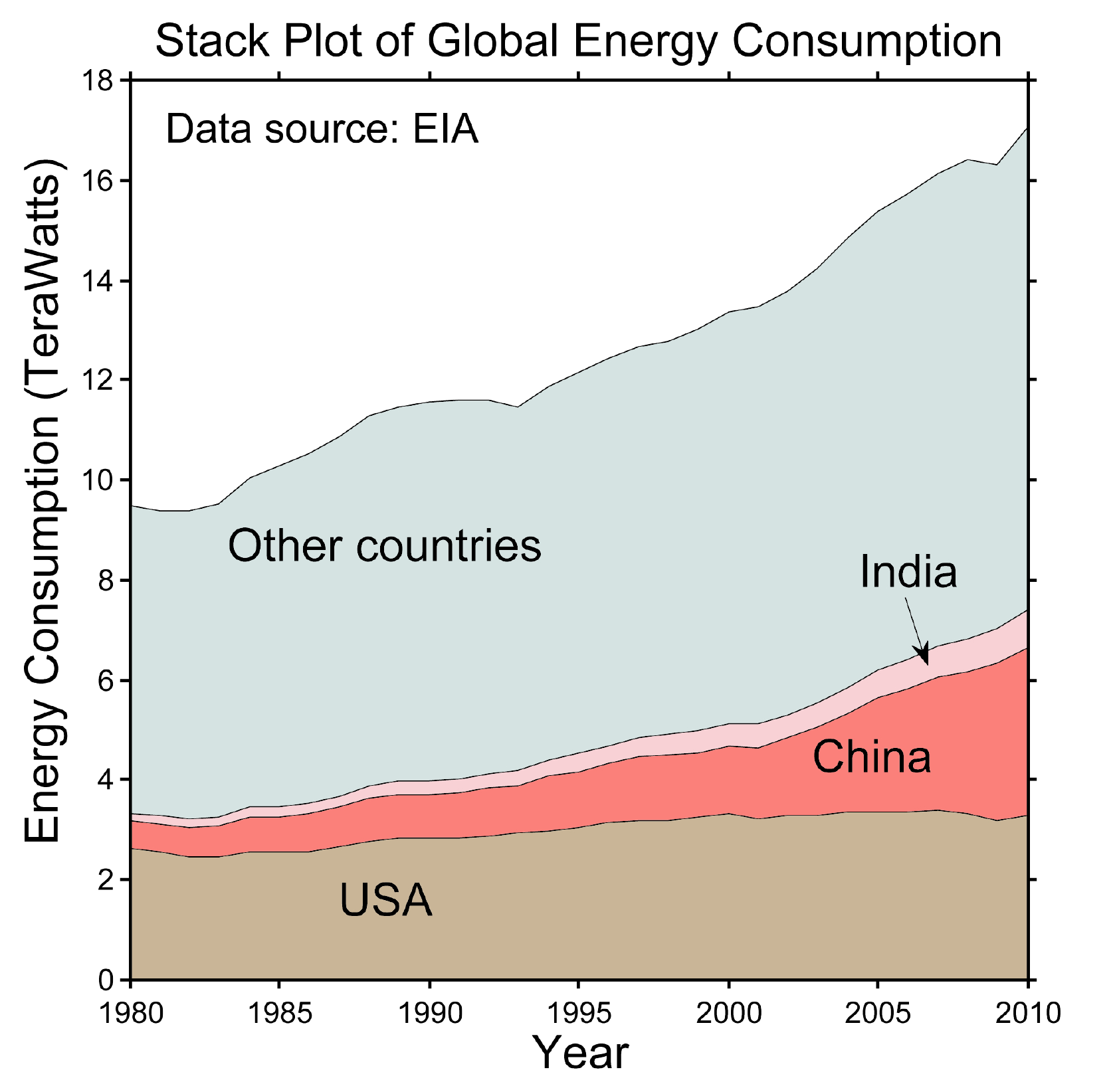} \hfill
\includegraphics[width=0.50\linewidth]{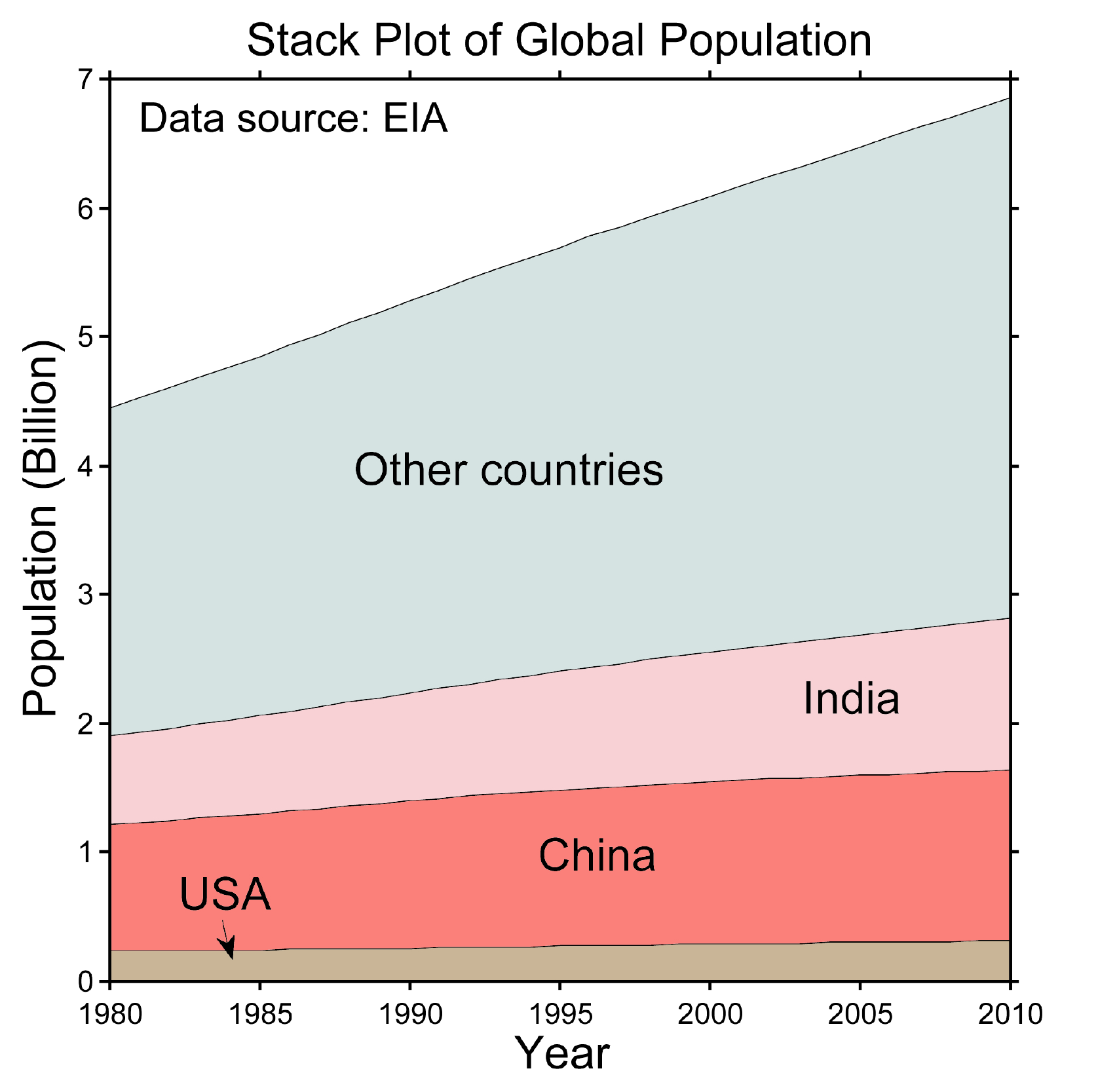}
\caption{Stack plots of energy consumption (\textbf{left panel}) and population (\textbf{right panel}) in USA, China, India, and the rest of the world from 1980--2010.}
\label{Fig:stacks}
\end{figure}

As energy consumption and population in developed countries have largely stabilized, most of the global increase comes from developing countries, such as China for energy (left panel in Fig.~\ref{Fig:stacks}) and India and others for population (right panel in Fig.~\ref{Fig:stacks}).  Thus, the issues of total energy consumption and carbon emissions are inextricably related to the problem of energy inequality among countries of the world.  Because countries have different sizes, it is appropriate to characterize each country $n$ by energy consumption per capita $\epsilon_n=E_n/N_n$, which, in thermodynamic terminology, is an intensive variable.  It is obtained by dividing the total energy consumption $E_n$ per unit time (i.e.\ the power) in each country $n$ by its population $N_n$, which are extensive variables in thermodynamic terminology.  Vast inequality in energy consumption per capita between developed and developing countries is generally well-recognized.  However, a global empirical probability distribution function $P_{\rm emp}(\epsilon)$ of energy consumption per capita $\epsilon$, based on data for all world countries, has not been constructed and studied quantitatively in the literature, to the best of our knowledge.  The main goal of our paper is to construct and interpret $P_{\rm emp}(\epsilon)$ using the EIA data \cite{EIA} for 1980--2010.  We believe that quantitative characterization of the global energy inequality and its historical trend is an important step toward understanding and solving challenging problems faced by society.

\section{Theoretical Analogy with Statistical Physics}
\label{Sec:theory}

Theoretical motivation for this study comes from an analogy with statistical physics, which was developed in previous papers by one of the authors \cite{Yakovenko-Web}.  For a gas in thermal equilibrium, the probability density function $P(\varepsilon)$ of finding an atom with the energy $\varepsilon$ is given by an exponential function\footnote{There is also a prefactor originating from the three-dimensional momentum space, but it is not relevant for our purposes, because we will only consider one-dimensional distributions.} of energy: $P_{\rm exp}(\varepsilon)\propto\exp(-\varepsilon/T)$, where $T$ is the temperature, and the Boltzmann constant is set to $k_B=1$.  This is the well-known Boltzmann--Gibbs probability distribution \cite{Wannier}.  There are few ``rich'' atoms with high energy and many ``poor'' atoms with low energy in this exponential distribution, so the energy distribution among the atoms is highly unequal, even though atoms (of the same kind) are identical.  This inequality is a consequence of the maximal entropy principle in the theory of probabilities \cite{Gorban}.  In accordance with the second law of thermodynamics, the equilibrium probability distribution $P_{\rm exp}(\varepsilon)$ is obtained by maximization of entropy subject to a constraint imposed by conservation of total energy in a closed system \cite{Wannier}.

In Ref.~\cite{Yakovenko-2000}, this general mathematical principle was applied to the probability distribution $P(m)$ of money $m$ in a statistical ensemble of economic agents engaging in monetary transactions.  In these transactions, which are similar to collisions between atoms in a gas, the agents transfer money among each other in payment for goods and services, but the total amount of money is conserved in a closed economic system.\footnote{Generally, only a central bank has the authority to issue new money.}  Under some restrictive conditions, a conjecture was made in Ref.~\cite{Yakovenko-2000} that the equilibrium probability distribution of money is exponential: $P_{\rm exp}(m)\propto\exp(-m/T_m)$, where $T_m=\langle m\rangle$ is the money temperature equal to the average amount of money per agent.\footnote{When debt is permitted, money balances can take negative values, which relaxes a boundary condition at $m=0$ and generally destabilizes the system \cite{Yakovenko-RMP,Yakovenko-2011}.}  Finding empirical statistical data for the distribution of money balances is difficult, so numerous papers subsequently focused on studying the empirical probability distributions of income in USA \cite{Yakovenko-2001a}, UK \cite{Yakovenko-2001b}, Australia \cite{Yakovenko-2006}, Romania \cite{Derzsy-2012}, European Union \cite{Jagielski-2013}, and other countries.  In most cases, a two-class structure of income distribution was found \cite{Yakovenko-2001b,Yakovenko-2003,Silva-2005,Banerjee-2010,Derzsy-2012,Jagielski-2013}, consisting of an exponential distribution for the majority of the population (about 97\%) in the lower class and a power law for the upper class (about 3\%).  More references can be found in the review article \cite{Yakovenko-RMP} and book \cite{Chakrabarti-2013}.

Besides income inequality within countries, there is also global inequality between rich and poor countries.  Characterizing global inequality by monetary description is difficult, because of different currencies with somewhat arbitrary and artificial exchange rates, although it is possible to use the purchasing power parity (PPP) instead \cite{Milanovic-2012}.  Banerjee and Yakovenko \cite{Banerjee-2010} took a different approach by focusing on inequality in energy consumption per capita $\epsilon$ measured in kilowatts (kW), which is a universal physical unit.  This variable is an indicator of physical standards of living in different countries and serves as a proxy for economic inequality around the world.  Treating global energy production (which predominantly comes from fossil fuels) as a limited resource\footnote{Strictly speaking, total energy production and consumption, as well as total population, do increase in time, as shown in Figs.~\ref{Fig:global-trends} and \ref{Fig:stacks}, but the rate of increase is relatively slow.  In this situation, entropy maximization subject to slowly changing constraints produces a quasi-equilibrium probability distribution with slowly changing parameters.} to be partitioned among the world population and applying the principle of maximal entropy, Ref.~\cite{Banerjee-2010} conjectured that the equilibrium probability distribution $P(\epsilon)$ of energy consumption per capita $\epsilon$ should be exponential: $P_{\rm exp}(\epsilon)\propto\exp(-\epsilon/T_\epsilon)$, where $T_\epsilon=\langle\epsilon\rangle=\int_0^\infty\epsilon\,P(\epsilon)\,d\epsilon$ is the average energy consumption per capita.

In order to verify this conjecture, Ref.~\cite{Banerjee-2010} studied a limited dataset downloaded from the World Resources Institute (WRI) covering about 130 countries for the period of 1990--2005 and found that the empirical probability distribution $P_{\rm emp}(\epsilon)$ evolves toward the exponential one $P_{\rm exp}(\epsilon)$.  However, some poorer countries with significant population, but low energy consumption, were missing from this dataset.  In the present paper, we use a much more detailed dataset from the EIA \cite{EIA} covering about 220 countries for the period of 1980--2010.  We find even stronger empirical support for the theoretical hypothesis proposed in Ref.~\cite{Banerjee-2010}.  We show that $P_{\rm emp}(\epsilon)$ has evolved from a highly unequal distribution in 1980 to a more equal distribution in 2010 that is quite close to exponential, as expected from the principle of maximal entropy.  We attribute this result to the globalization of the world economy, which mixes the world and brings it closer to the state of maximal entropy.

\newpage
\section{Energy Consumption Distribution}
\label{sec:energy}
\vspace{-12pt}
\subsection{Cumulative Probability Distribution Function}
\label{sec:cumulative}

For a discrete number of about 220 countries in the EIA dataset \cite{EIA}, it is more convenient to construct the empirical complementary cumulative distribution function (CDF) for energy consumption defined as $C_{\rm emp}(\epsilon)=\int_\epsilon^\infty P_{\rm emp}(\epsilon')\,d\epsilon'$, rather than the probability density function (PDF) $P_{\rm emp}(\epsilon)$.  First, we sort all countries in the ascending order of their energy consumption per capita $\epsilon_n$, so that $n=1$ corresponds to the country with the lowest consumption and $n=L$ to the maximal consumption, where $L$ is the total number of countries.  Then, the cumulative distribution function is 
  \begin{equation}  \label{C(e_n)}  
  C_{\rm emp}(\epsilon_n) = \frac{\sum_{k=n+1}^L N_k}{\sum_{k=1}^L N_k},  
  \end{equation}
which is the fraction of world population whose energy consumption per capita is greater than $\epsilon_n$.  Effectively, this construction assigns the same energy consumption $\epsilon_n=E_n/N_n$ to all $N_n$ residents of the country $n$.  This approach corresponds to Concept 2 in studies of global inequality, according to the terminology from the book \cite{Milanovic-apart}.  In Concept 1, equal weights are assigned to all countries irrespective of their populations, so a group of tiny countries can outweigh the few most populous countries in the distribution of $\epsilon_n$.  A more sensible approach is to assign weights to different countries proportional to their populations, which is done in Concept 2 and in the CDF in Eq.~(\ref{C(e_n)}).  The more sophisticated Concept 3 takes into account probability distributions (i.e.~inequality) within countries and combines them into a global probability distribution.  Although Concept 3 is the most accurate, it is very difficult to find the required data, so we only use Concept 2 in our paper.

\begin{figure}
\includegraphics[width=0.50\linewidth]{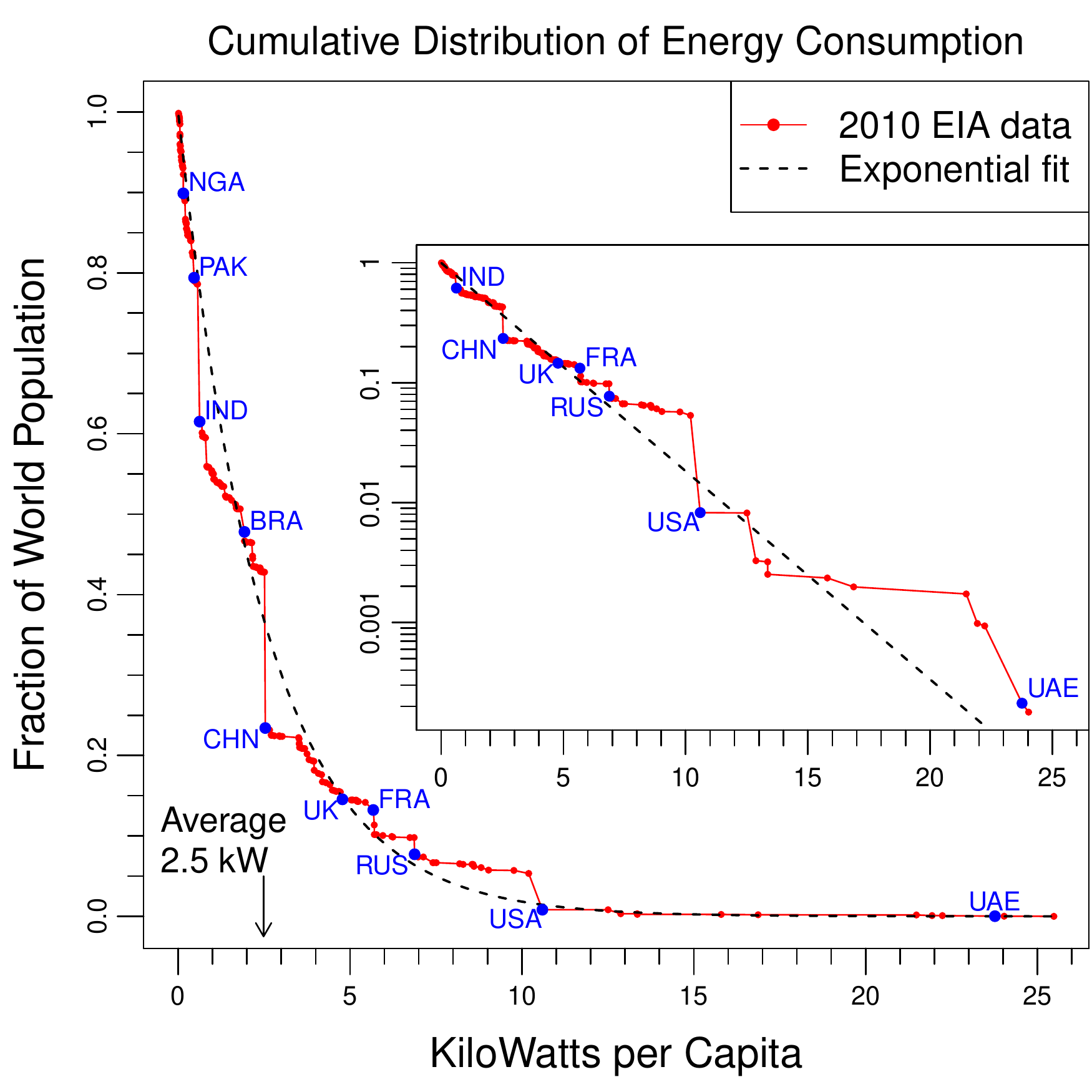} \hfill
\includegraphics[width=0.48\linewidth]{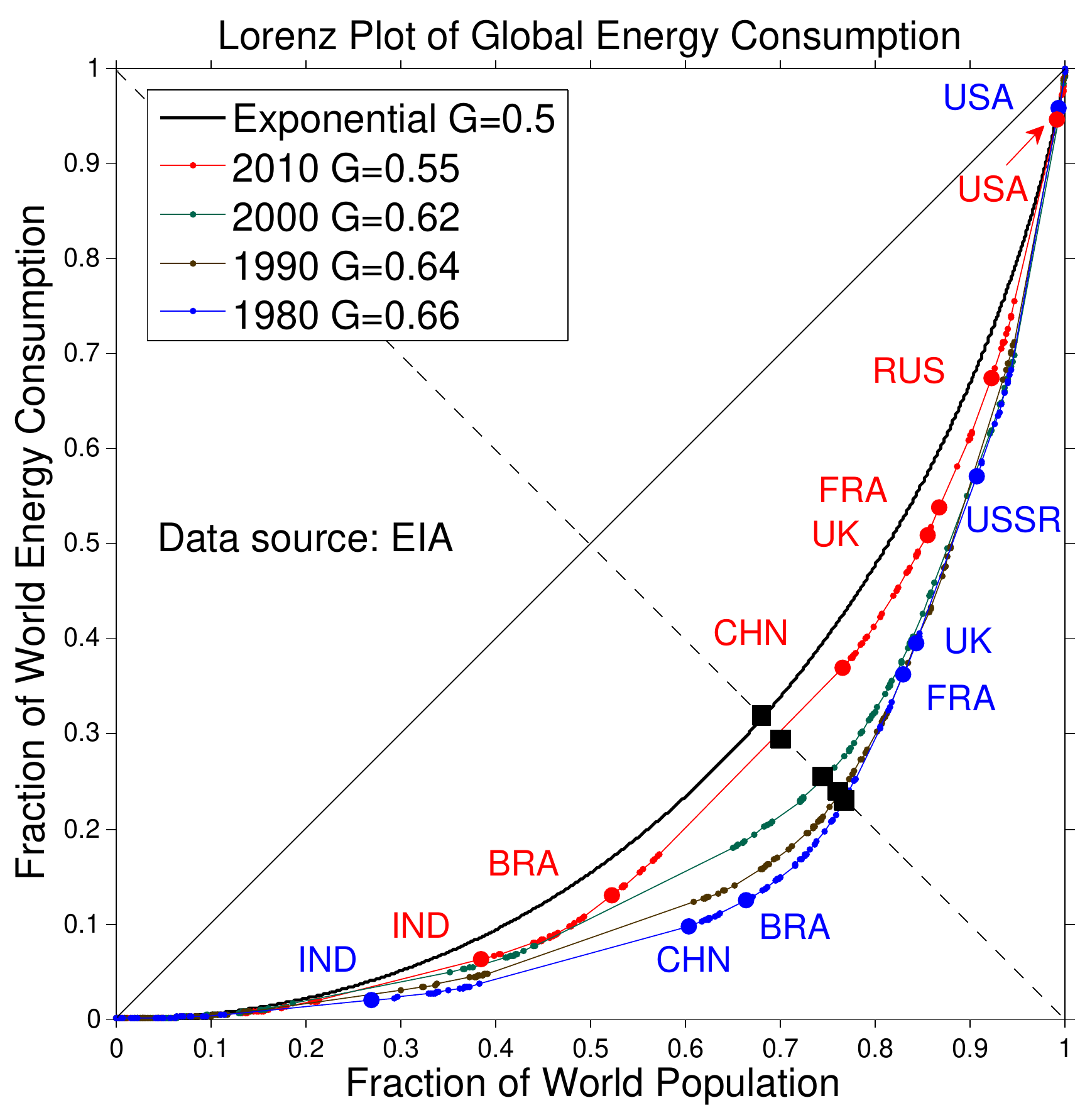}
\caption{(\textbf{Left panel}) Complementary cumulative probability distribution function, Eq.~(\ref{C(e_n)}), of the global energy consumption per capita in 2010 (red curve), compared with an exponential fit (black dashed curve).  The inset shows the same plot in log-linear scale.  (\textbf{Right panel}) Lorenz plots for the global energy consumption per capita in 1980--2010 (colored curves), compared with an exponential distribution (black solid curve).}
\label{Fig:CDF-energy}
\end{figure}

The empirical CDF $C_{\rm emp}(\varepsilon_n)$ constructed from the EIA data for 2010 is shown in the left panel of Fig.~\ref{Fig:CDF-energy} by the red solid line with circles.  The global average energy consumption per capita is
  \begin{equation}  \label{<e>}  
  \langle\epsilon\rangle=\frac{\sum_{k=1}^L \epsilon_n N_k}{\sum_{k=1}^L N_k}=\frac{E}{N}
  \approx 2.5~\mbox{kW in 2010},
  \end{equation}
where $E$ is the total global energy consumption, and $N$ is the total global population.  In Fig.~\ref{Fig:CDF-energy},  we observe broad inequality, where $\epsilon=10$ kW in the USA is 4 times \textit{greater} than $\langle\epsilon\rangle$, $\epsilon=0.6$ kW in India is 4 times \textit{lower} than $\langle\epsilon\rangle$, and $\epsilon=2.5$ kW in China is approximately equal to $\langle\epsilon\rangle$.

The black dashed line in the left panel of Fig.~\ref{Fig:CDF-energy} shows the theoretical exponential probability distribution $C_{\rm exp}(\epsilon)=\exp(-\epsilon/T_\epsilon)$, which gives a reasonably good fit of the whole distribution using $T_\epsilon=\langle\epsilon\rangle=2.5$~kW.  The inset shows a log-linear plot of $C_{\rm emp}(\varepsilon_n)$, where the vertical axis is logarithmic and the horizontal axis is linear.  In these coordinates, the exponential function becomes the straight dashed line, and the red empirical points fall reasonably close to the theoretical line.  Thus, we conclude that the empirical probability distribution of energy consumption per capita in 2010 is close to exponential.

\newpage
\subsection{Lorenz Curves}
\label{sec:Lorenz}

To put this observation in historical perspective, we construct the Lorenz curves for energy consumption per capita from 1980--2010.  For a given probability density $P(\epsilon)$, let us introduce the variables
  \begin{equation}
  x(\epsilon)=1-C(\epsilon)=\int_0^\epsilon P(\epsilon')\,d\epsilon', \qquad
  y(\epsilon)=\frac{1}{\langle\epsilon\rangle}
  \int_0^\epsilon \epsilon\,P(\epsilon')\,d\epsilon',
  \label{xy(e)}
  \end{equation}
where $x(\epsilon)$ is the fraction of the population whose energy consumption per capita is below $\epsilon$, and $y(\epsilon)$ is the total energy consumption of this population normalized by $\langle\epsilon\rangle$.  The Lorenz curve \cite{Lorenz} is a parametric plot of $y(\epsilon)$ versus $x(\epsilon)$, where the parameter $\epsilon$ varies from 0 to $\infty$.    The variables $x$ and $y$ are bounded between 0 and 1, so the Lorenz curve connects the points (0,0) and (1,1) in the $(x,y)$ plane.

To construct a Lorenz curve from empirical data, we calculate $x_{\rm emp}(\epsilon_n)$ and $y_{\rm emp}(\epsilon_n)$ for a set of countries ordered by the index $n$ from the lowest to the highest energy consumption per capita $\epsilon_n$:
  \begin{equation}
  x_{\rm emp}(\epsilon_n) = 1-C_{\rm emp}(\epsilon_n) = \frac{\sum_{k=1}^n N_k}{\sum_{k=1}^L N_k}, \qquad
  y_{\rm emp}(\epsilon_n) = \frac{\sum_{k=1}^n \epsilon_k N_k}{\sum_{k=1}^L \epsilon_k N_k}.
  \label{xy(e_n)}  
  \end{equation}
The right panel in Fig.~\ref{Fig:CDF-energy} shows the parametric Lorenz plots of $y_{\rm emp}(\epsilon_n)$ versus $x_{\rm emp}(\epsilon_n)$ constructed from the EIA data for 1980, 1990, 2000, and 2010.  Over this time period, the Lorenz plots have moved up, which indicates that inequality in energy consumption per capita has decreased.  However, even the latest Lorenz curve for 2010 is still very far from the straight diagonal line, which would correspond to perfect equality where all countries would have equal $\epsilon_n$.

The Lorenz curve for the theoretically expected exponential probability distribution discussed in Sec.~\ref{Sec:theory} was calculated in Ref.~\cite{Yakovenko-2001a} by substituting $P_{\rm exp}(\epsilon)$ into Eq.~(\ref{xy(e)}) and eliminating $\epsilon$:
  \begin{equation}  \label{y(x)}
  y_{\rm exp}=x+(1-x)\ln(1-x).
  \end{equation}
This function, which has no fitting parameters, is shown by the solid black curve in the right panel of Fig.~\ref{Fig:CDF-energy}.  We observe that the empirical red Lorenz curve for 2010 is quite close to the theoretical black curve derived for the exponential distribution.  The full time evolution of the Lorenz curves from 1980 to 2010 is shown in a computer animation movie in the supplementary online material.  The movie shows the Lorenz curve evolving from a highly unequal distribution in 1980 to a more equal distribution in 2010 that is quite close to the exponential distribution expected from the principle of maximal entropy.  This observation is in qualitative agreement with the earlier results presented in Ref.\ \cite{Banerjee-2010} for a more limited dataset for 1990--2005.

\begin{figure}[H]
\includegraphics[width=0.49\linewidth]{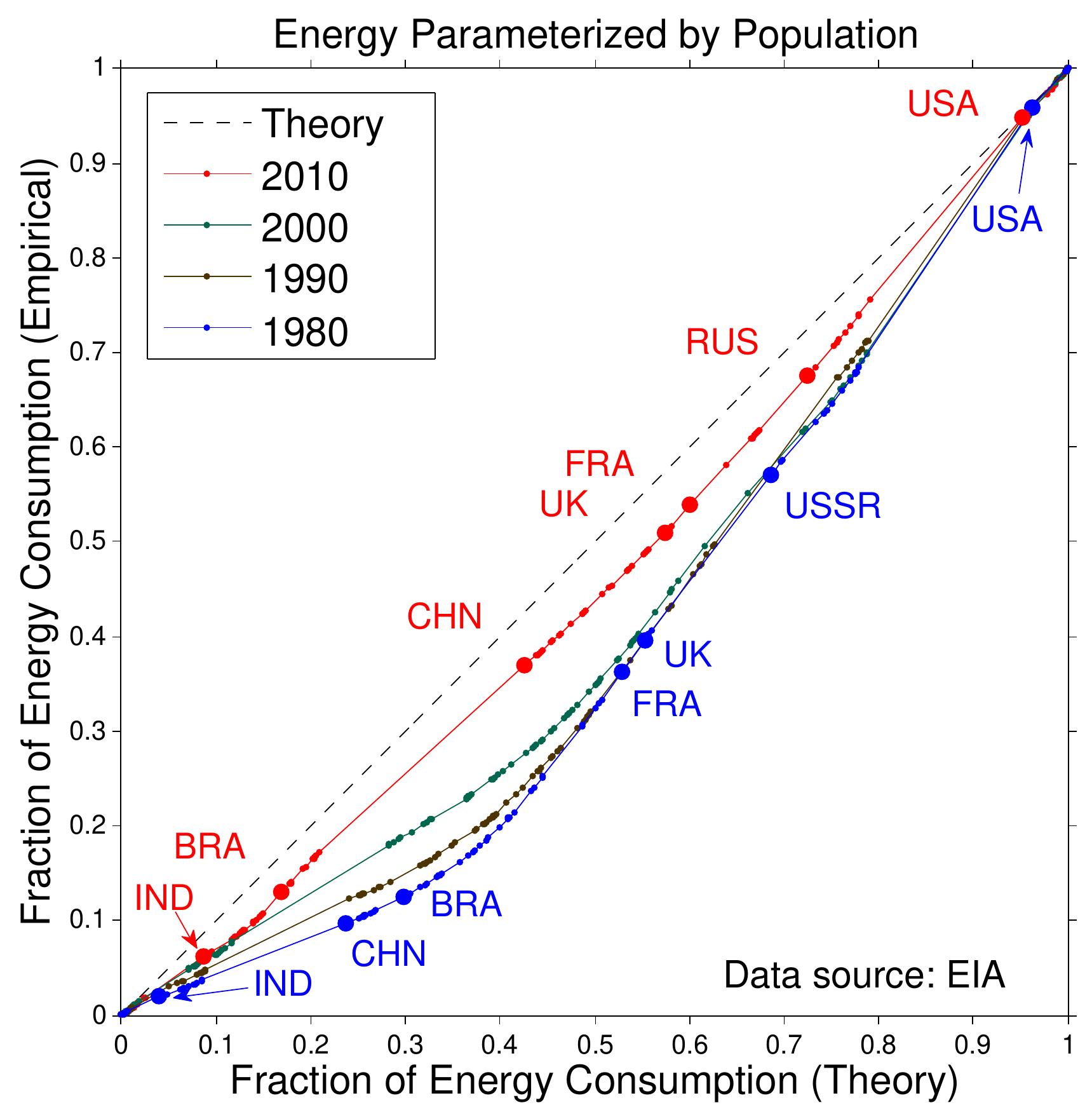} \hfill
\includegraphics[width=0.49\linewidth]{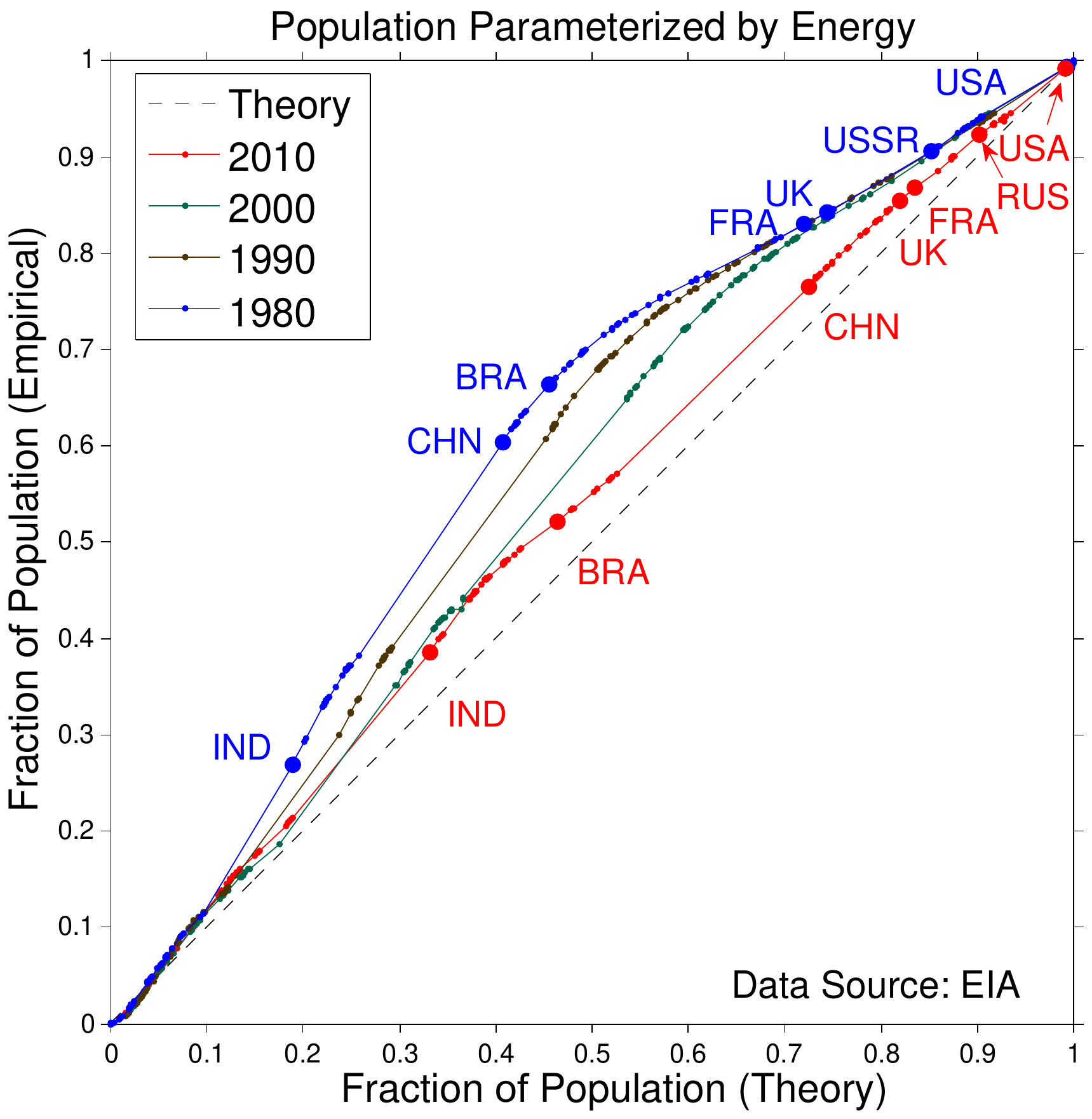}
\caption{(\textbf{Left panel}) Parametric plots of the empirical $y_{\rm emp}(x)$ versus exponential $y_{\rm exp}(x)$ cumulative fractions of global energy consumption using the population fraction $x$ as a parameter.  (\textbf{Right panel}) Parametric plots of the empirical $x_{\rm emp}(y)$ versus exponential $x_{\rm exp}(y)$ cumulative population fractions using the energy consumption fraction $y$ as a parameter.}
\label{Fig:PP-plots}
\end{figure}

Convergence of the empirical Lorenz curve $y_{\rm emp}(x)$, Eq.~(\ref{xy(e_n)}), toward the theoretical one $y_{\rm exp}(x)$, Eq.~(\ref{y(x)}), is also illustrated by the parametric plots in Fig.~\ref{Fig:PP-plots}.  The left panel shows the plots of $y_{\rm emp}(x)$ versus $y_{\rm exp}(x)$ using $x$ as a parameter, whereas the right panel shows the plots of $x_{\rm emp}(y)$ versus $x_{\rm exp}(y)$ using $y$ as a parameter.  In both panels, as time progresses from 1980 to 2010, the curves move toward the dashed diagonal line corresponding to perfect agreement between empirical and theoretical Lorenz curves.

\newpage
\subsection{Gini Coefficient}
\label{sec:Gini}

A standard measure of inequality is the Gini coefficient $G$.  It is defined as the area between the Lorenz curve and the solid diagonal line in the right panel of Fig.~\ref{Fig:CDF-energy}, divided by the area 1/2 of the triangle beneath the diagonal line.  The Gini coefficient $0\leq G\leq 1$ varies from 0 at perfect equality, where everybody receives an equal share, to 1 at extreme inequality, where everybody receives nothing, except for one person who receives everything.  It was shown in Ref.~\cite{Yakovenko-2001a} that $G=0.5$ for an exponential distribution.  

\begin{figure}[H]
\includegraphics[width=0.49\linewidth]{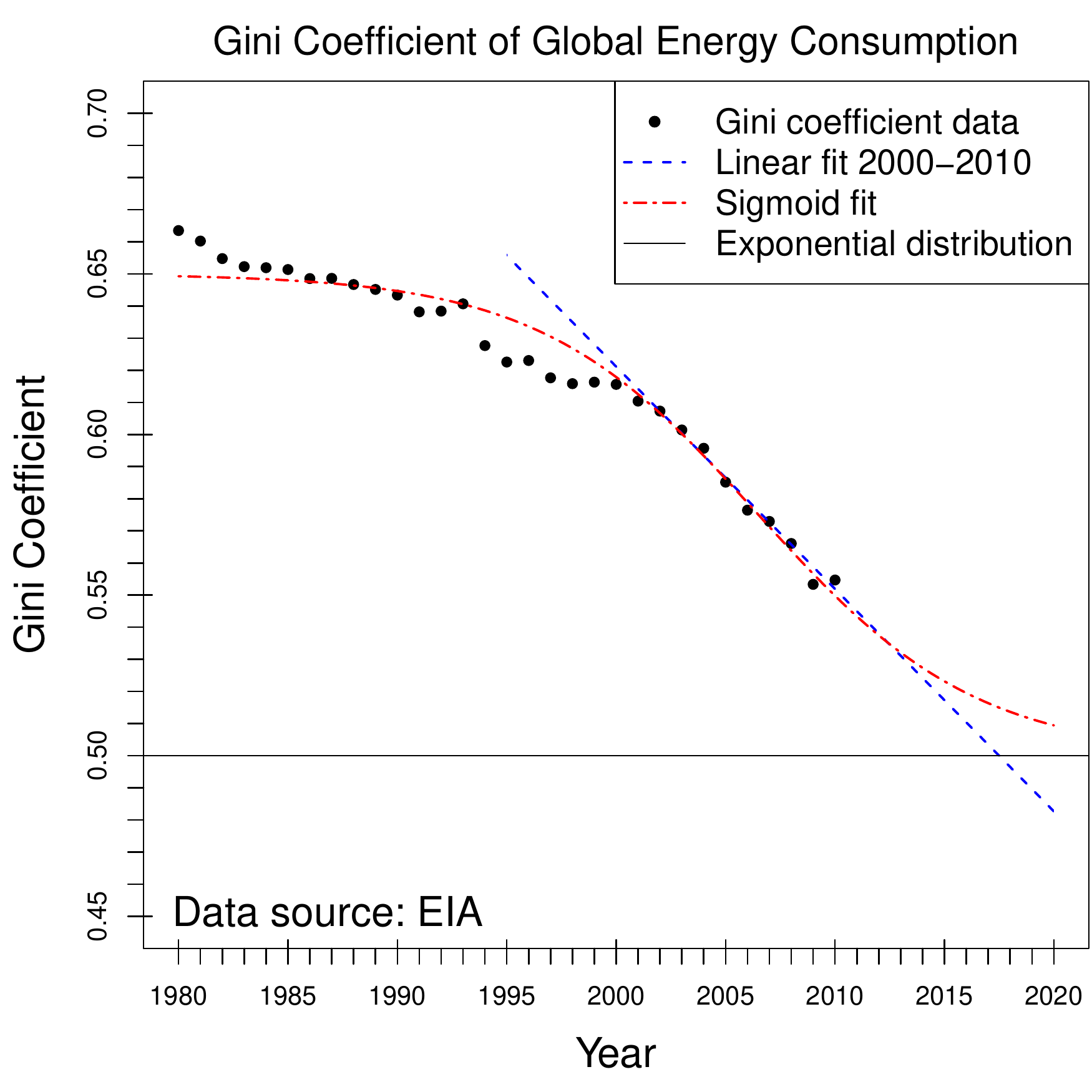} \hfill
\includegraphics[width=0.49\linewidth]{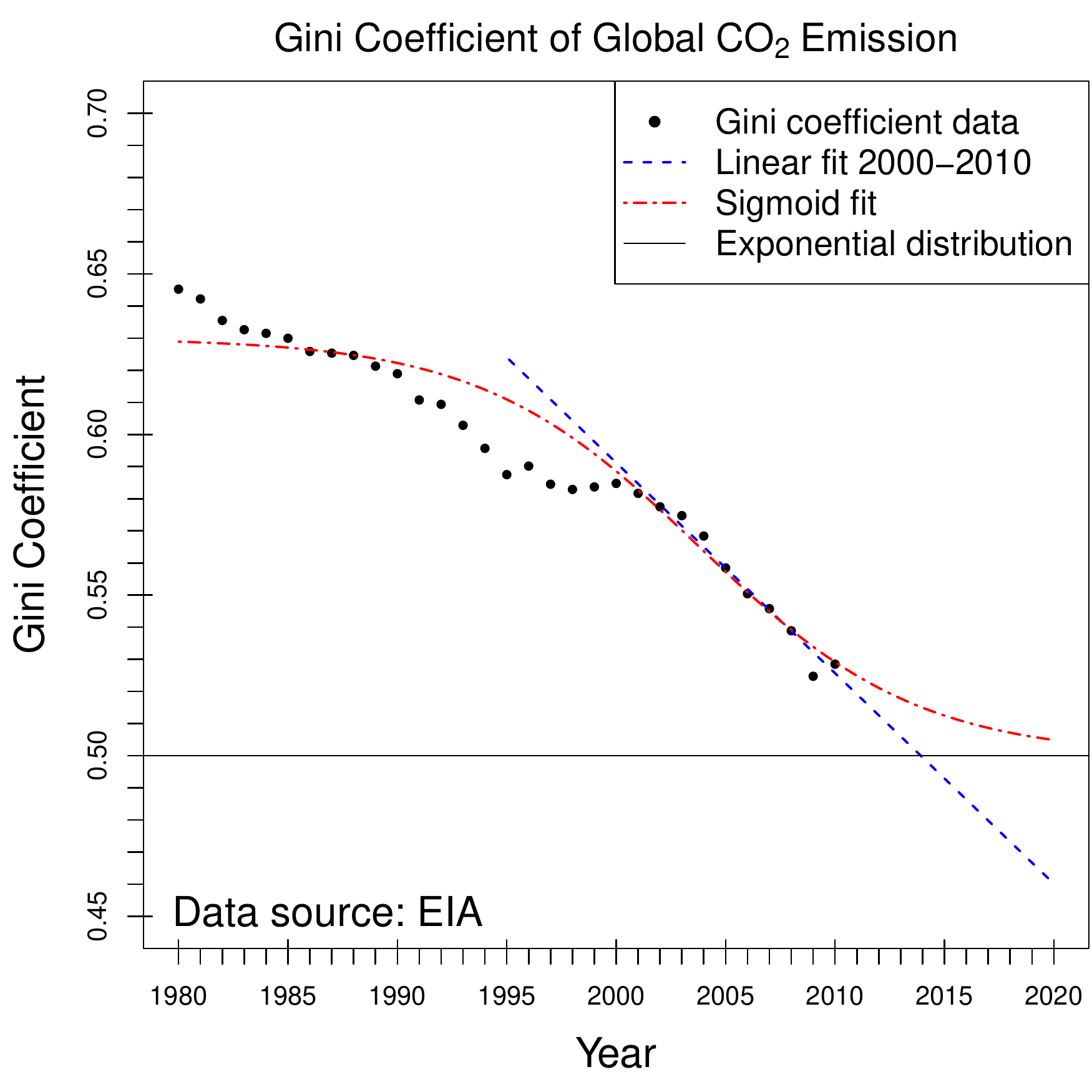}
\caption{Historical evolution of the global Gini coefficient $G$ (circles) for energy consumption per capita (\textbf{left panel}) and $\rm CO_2$ emissions per capita (\textbf{right panel}).  The blue dashed line represents a linear extrapolation, and the red dashed-dotted curve a sigmoid fit.  The horizontal line at $G=0.5$ corresponds to an exponential distribution.}
\label{Fig:Gini}
\end{figure}

The inset in the right panel of Fig.~\ref{Fig:CDF-energy} shows that the Gini coefficient for energy consumption per capita has decreased from 0.66 to 0.55 during 1980--2010 and now is close to the theoretical equilibrium value of 0.5.  The full historical evolution of the Gini coefficient is shown in the left panel of Fig.\ \ref{Fig:Gini} by the circles, whereas the horizontal line indicates the theoretical equilibrium value $G=0.5$ for an exponential distribution.  We observe that the Gini coefficient was decreasing particularly fast in the last decade of 2000--2010.  We attribute this effect to rapid globalization of the world economy, which brings the world closer to the state of maximal entropy.  By making a naive linear extrapolation of the trend for the last 10 years, shown by the blue dashed line in Fig.\ \ref{Fig:Gini}, one might expect that the Gini coefficient will reach $G=0.5$ around 2017 and will continue to decrease toward perfect equality.  In contrast, the principle of maximal entropy suggests that the decrease of energy consumption inequality would slow down and saturate at the level $G=0.5$ corresponding to the exponential distribution.  Thus, we make a conjecture that the Gini coefficient will follow a sigmoid curve, shown by the red dashed-dotted line in Fig.~\ref{Fig:Gini}, asymptotically approaching $G=0.5$, but not going below this value.  We plan to revisit this study in five and ten years from now, when new data will become available, and verify our prediction.  Making predictions about the future is always a challenging task.  In physics, theories that not only explain known experiments, but also make successful predictions about future observations are particularly valuable.  Our conjecture is, in principle, falsifiable, i.e.\ it may be proven wrong by future observations. 

There is widespread discussion in the media about coming global economic slowdown, which is sometimes called the ``economic ice age'' \cite{Dent,Ice-Age,India}.  In developed countries, the population has been aging and consumption (including energy consumption) has been decreasing for some time already, accompanied by stagnation of economic growth.  In contrast, economic growth (and the corresponding growth of energy consumption) was fast in developing countries, particularly in China, in the last few decades.  However, there are indications that this economic growth is ending too.  Commentators offer various specific reasons for the global economic slowdown \cite{Ice-Age}.  The export-oriented growth model is unsustainable, because it requires ever increasing consumption in developed countries, which is not the case any more.  The population of China has stabilized and is beginning to age.  India is mired in rupee inflation, underdeveloped infrastructure, and gloomy economic prospects \cite{India}.  However, in our opinion, the underlying reason for various manifestations of the global economic slowdown is actually entropic \cite{Yakovenko-2013}.  As shown in the above graphs, the energy consumption inequality in 1980 was much higher than expected for thermal statistical equilibrium, so the entropy of the distribution was lower.  This deviation from statistical equilibrium was driving globalization, thus increasing consumption in developing countries and decreasing global inequality.  But now, the world is close to the state of maximal entropy and thermal statistical equilibrium, so the driver for further evolution disappears, and the world is likely to stay put in the present state of global inequality.  This reasoning is somewhat similar to the argument for the ``thermal death of the Universe'' much discussed in 19th century physics.  In the 20th century, it was recognized that one way around this argument is expansion of the Universe.  Similarly, human development for centuries was driven by geographic expansion, but this era is over, and now the world is small, hot, and ``flat'' \cite{Friedman-flat}.

However, the ``economic ice age'' may have the beneficial effect of slowing down climate change.  The EIA projections in Fig.~\ref{Fig:global-trends} are based on the assumption of accelerated global economic growth in the future.  The blue circles and crosses show an increase in global energy consumption and $\rm CO_2$ emissions per capita in 2020--2040, departing from the relatively stable level in 1980--2010.  In contrast, we expect that per-capita levels would stay approximately constant for the global slowdown scenario, so the total energy consumption and $\rm CO_2$ emissions (red circles and crosses) would only increase in proportion to the population growth (black curve), as they did for most of 1980--2010.

\subsection{The Law of 1/3}
\label{sec:1/3}

Given that the distribution of energy consumption per capita is already close to exponential, here we formulate a new and important law, which we call the law of 1/3.  It is analogous to the well-known Pareto's principle of 80--20.  Pareto observed in 1906 that 80\% of land in Italy was owned by 20\% of the people \cite{Pareto-20-80}.  Conversely, 80\% of the people owned only 20\% of the land. Since then, it was claimed by many authors that this principle applies to various other statistical distributions, although it is not always clear whether these claims are based on actual data or are just rhetorical flourishes.  For any probability distribution of resource ownership, it is always possible to find a fraction $x^*$ of population owning the $1-x^*$ fraction of the resource, e.g.\ $x^*=0.8$ for Pareto's principle.  Mathematically, the value of $x^*$ is obtained by solving the equation $y(x^*)=1-x^*$, where $y(x)$ is the Lorenz curve.  Geometrically, it is obtained as an intersection of the Lorenz curve and the dashed diagonal line in the right panel of Fig.\ \ref{Fig:CDF-energy}, as shown by the squares.  Using Eq.~(\ref{y(x)}) for an exponential distribution, we find $x_{\rm exp}^*\approx0.68\approx2/3$ from the equation $1-x^*=x^*+(1-x^*)\ln(1-x^*)$, whereas the empirical value for the 2010 data is $x_{\rm emp}^*\approx0.69$.  Since these values are quite close, we can say that \textbf{the top 1/3 of the world population consumes 2/3 of energy}, which we call \textbf{the law of 1/3}.  This simple and easily understandable statement summarizes the current state of global inequality in energy consumption.  To our knowledge, this result has not been noted before.  Because it is a consequence of the principle of maximal entropy, we expect that the global energy consumption inequality will stay at this level in the future.

\begin{figure}[H]
\includegraphics[width=0.50\linewidth]{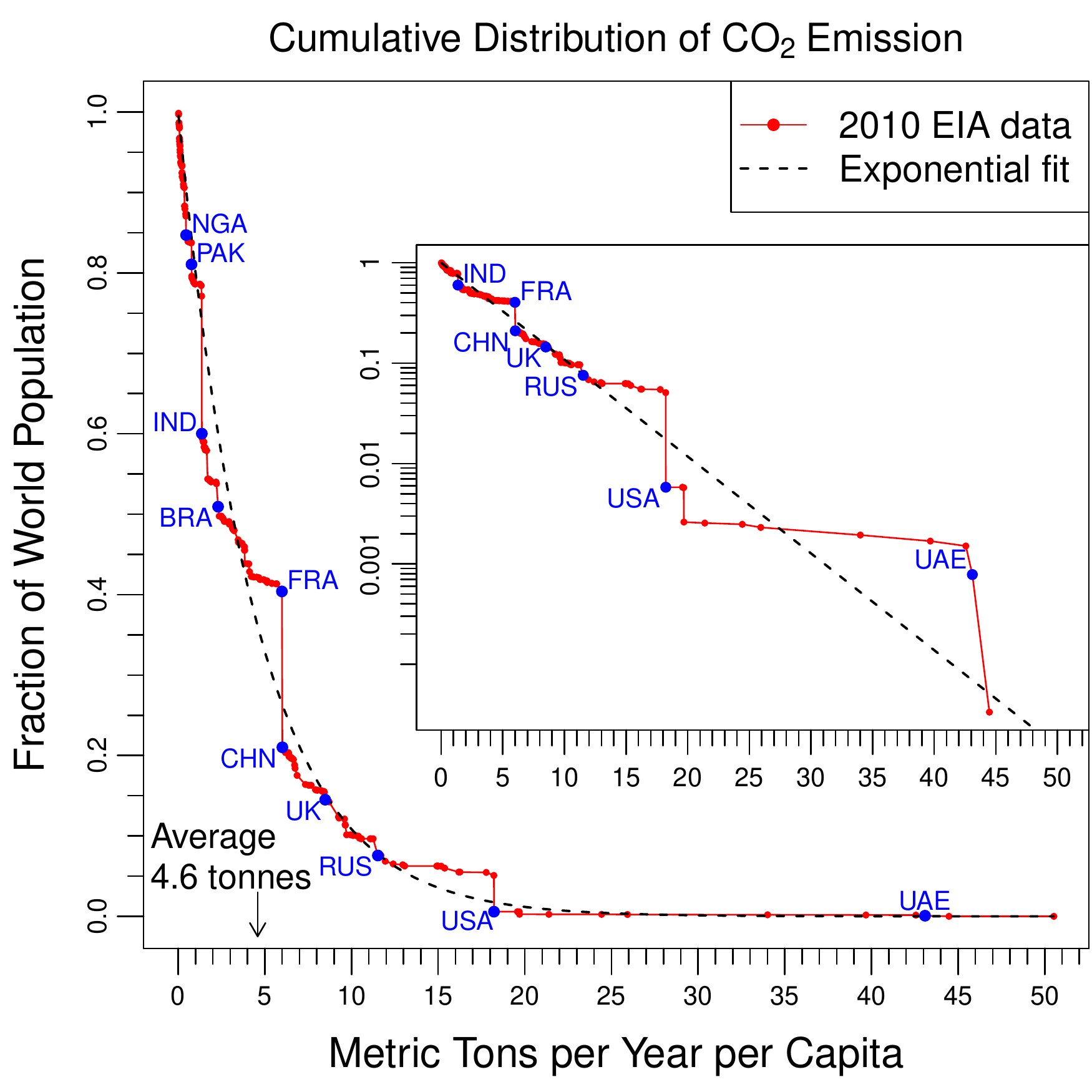} \hfill
\includegraphics[width=0.49\linewidth]{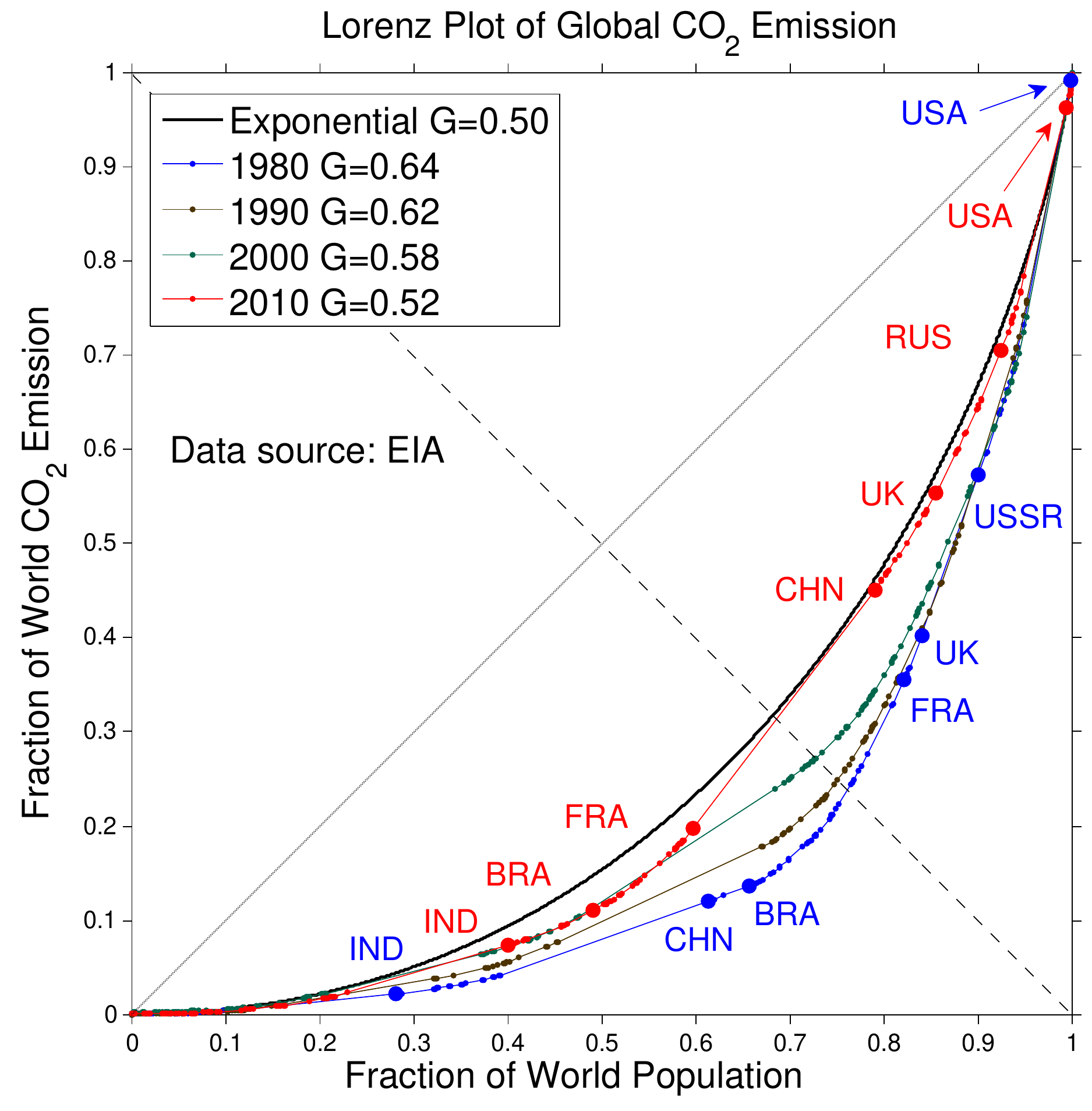}
\caption{(\textbf{Left panel}) Complementary cumulative probability distribution function of the global $\rm CO_2$ emissions per capita in 2010 (red curve), compared with an exponential fit (black dashed curve).  The inset shows the same plot in log-linear scale.  (\textbf{Right panel}) Lorenz plots for the global $\rm CO_2$ emissions per capita in 1980--2010 (colored curves), compared with an exponential distribution (black solid curve).}
\label{Fig:CDF-CO2}
\end{figure}

\section{$\mathbf{CO_2}$ Emissions Distribution}
\label{Sec:CO2}

\begin{figure}[H]
\includegraphics[width=0.49\linewidth]{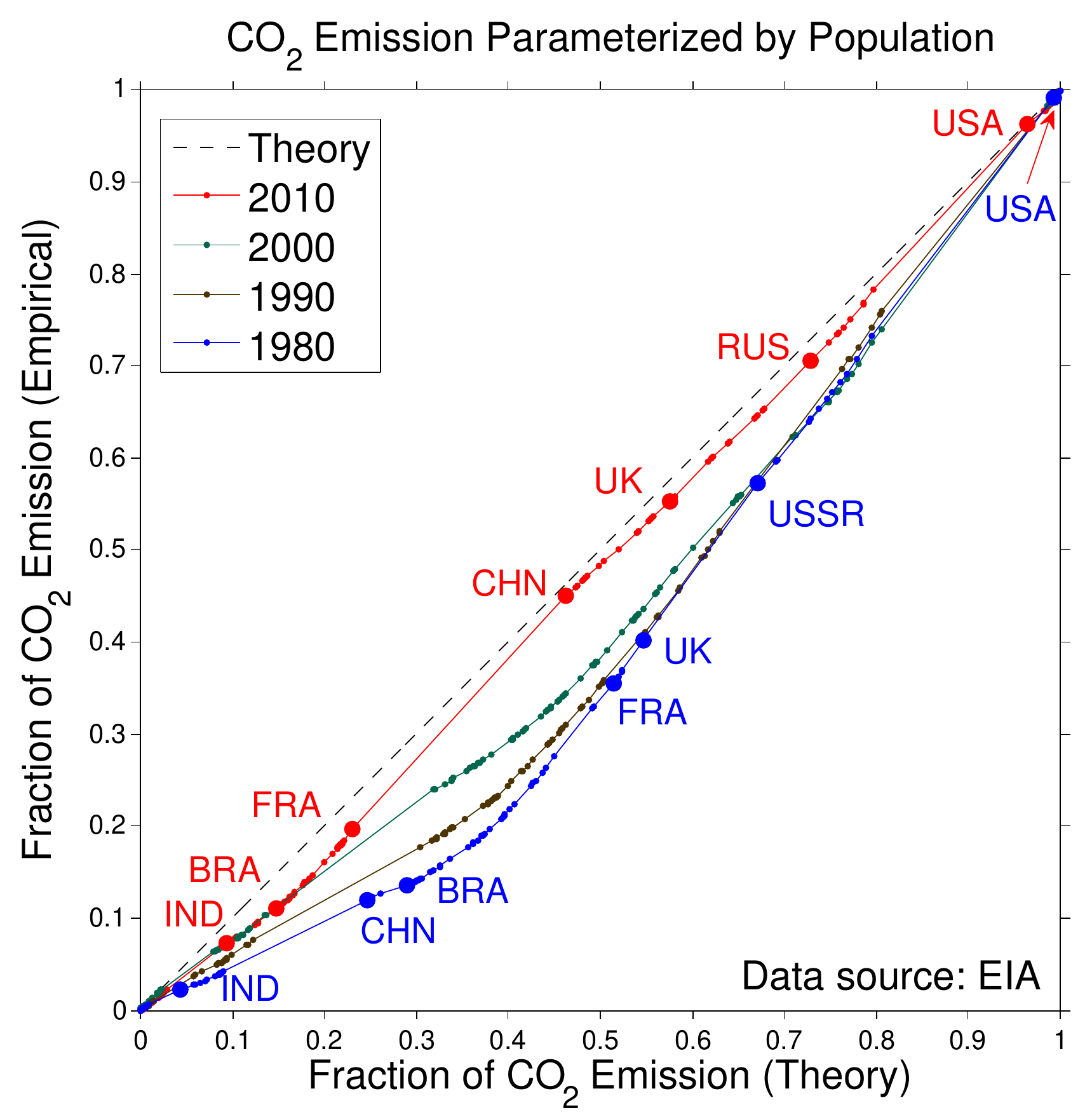} \hfill
\includegraphics[width=0.49\linewidth]{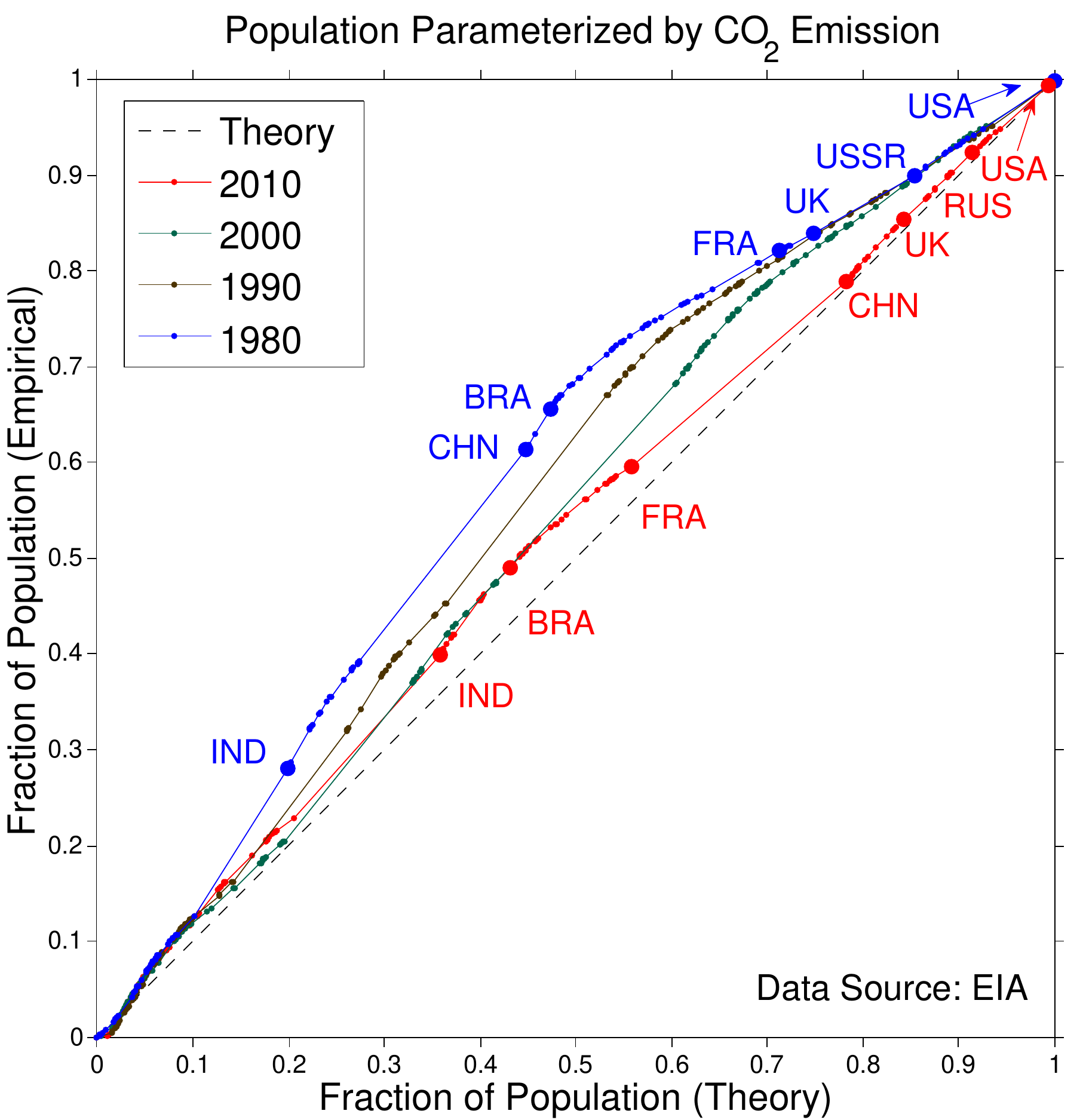}
\caption{(\textbf{Left panel}) Parametric plots of the empirical $y_{\rm emp}(x)$ versus exponential $y_{\rm exp}(x)$ cumulative fractions of global $\rm CO_2$ emissions using the population fraction $x$ as a parameter.  (\textbf{Right panel}) Parametric plots of the empirical $x_{\rm emp}(y)$ versus exponential $x_{\rm exp}(y)$ cumulative population fractions using the $\rm CO_2$ emissions fraction $y$ as a parameter.}
\label{Fig:QQ-CO2}
\end{figure}

Here we present a study of global inequality in $\rm CO_2$ emissions per capita, for which we follow the same procedure as in Sec.~\ref{sec:energy}, but for carbon emissions instead of energy consumption.  The resulting complementary cumulative distribution function for 2010 and the Lorenz curves for 1980--2010 are shown in the left and right panels of Fig.~\ref{Fig:CDF-CO2} respectively.  The full time evolution of the Lorenz curves for carbon emissions from 1980 to 2010 is shown in a computer animation movie in the supplementary online material.  Again, we observe that the global distribution of carbon emissions per capita is quite close to exponential in 2010.  The corresponding parametric plots for carbon emissions are shown in Fig.~\ref{Fig:QQ-CO2} in a manner similar to Fig.~\ref{Fig:PP-plots}.  Interestingly, the global distribution of carbon emissions per capita in 2010 is even closer to exponential than the distribution of energy consumption per capita discussed in Sec.~\ref{sec:energy}.  Historical evolution of the Gini coefficient for carbon emissions is shown in the right panel of Fig.~\ref{Fig:Gini}, in comparison with the Gini coefficient for energy consumption shown in the left panel.  The plots in the left and right panels look quite similar.  Moreover, they are also similar to the plot of the Gini coefficient for global \emph{income} inequality shown in Fig.~1 of Ref.~\cite{Milanovic-2012}, where it was constructed using Concept 2.  The similarity of all three Gini plots indicates that energy consumption, carbon emissions from fossil fuels, and monetary income are, essentially, interchangeable measures of global inequality, because each of these measures currently reflects the standards of living around the world.

The similarity of the distributions for energy consumption and carbon emissions per capita is not surprising, because most of the world energy consumption is derived from carbon-intensive fossil fuels.  However, the carbon intensities of different types of fossil fuels, such as coal, petroleum, and natural gas, are somewhat different.  Apparently, this difference averages out, so that the global distributions of energy consumption and carbon emissions are quite similar.  For example, the exceptional position of France (FRA) does not affect the overall global distributions significantly.  Because most of its electricity is generated from nuclear energy, France has relatively high energy consumption shown in the right panel of Fig.~\ref{Fig:CDF-energy} and relatively low carbon emissions shown in the right panel of Fig.~\ref{Fig:CDF-CO2}, but the two global distributions nevertheless remain quite similar.

\begin{figure}[H]
\centerline{\includegraphics[width=0.5\linewidth]{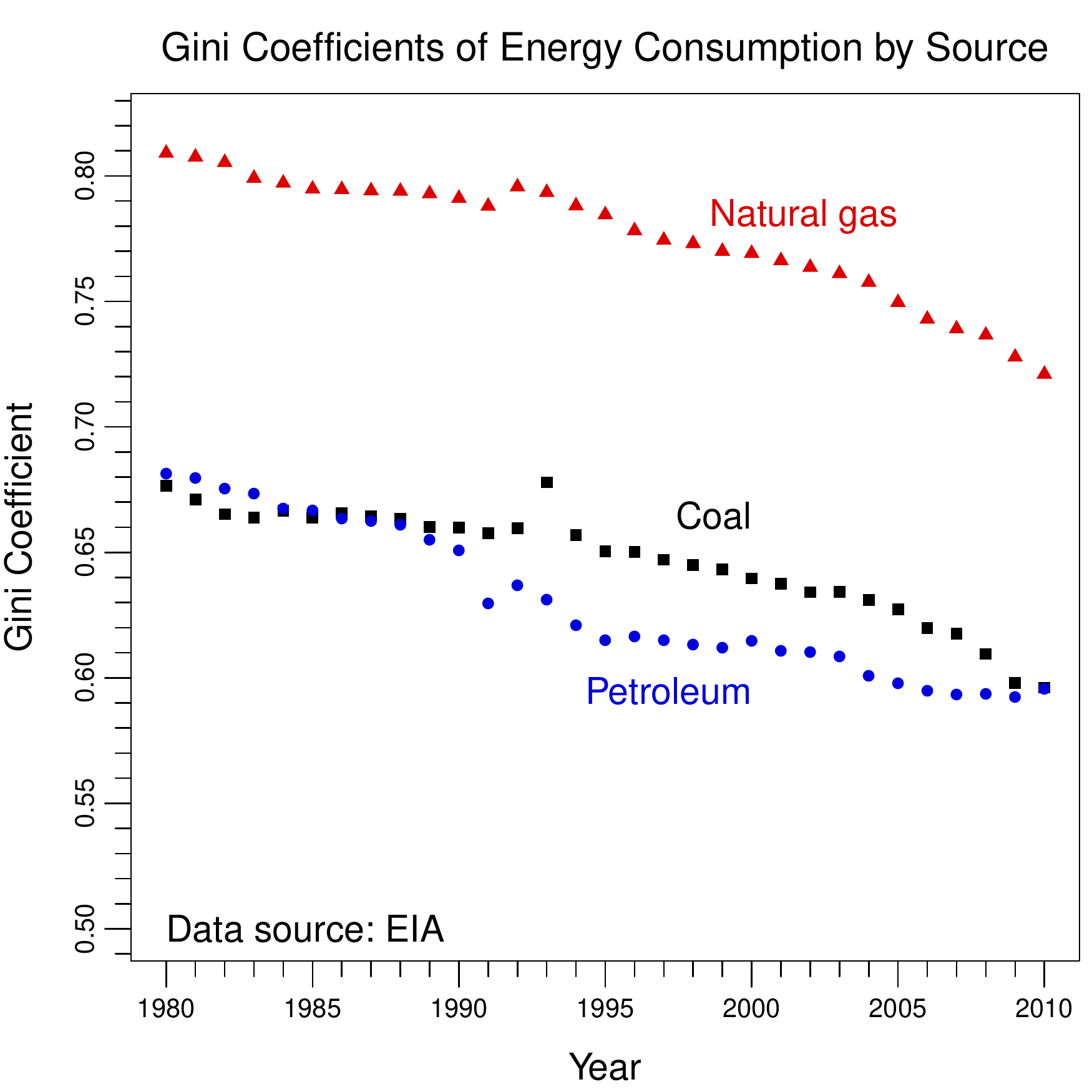}}
\caption{Historical evolution of the global Gini coefficients for natural gas, coal, and petroleum  consumption per capita.}
\label{Fig:Gini-by-source}
\end{figure}

The underlying assumption for the application of the principle of maximal entropy in our paper is that energy resources (typically fossil fuels) are extracted in some places (e.g.\ in the Persian Gulf) and are redistributed for consumption in different places (e.g.\ in the USA).  Petroleum and coal are relatively easy to transport by sea, but transcontinental transportation of liquified natural gas is more difficult.  Thus, we would expect a higher Gini coefficient for consumption of natural gas per capita compared with coal and petroleum, because natural gas is more difficult to transport on a global scale from the places of its natural abundance.  Figure \ref{Fig:Gini-by-source} shows that this is indeed the case.  While the Gini coefficients for coal and petroleum consumption per capita are relatively close, the Gini coefficient for natural gas consumption per capita is much higher.  Nevertheless, all three Gini coefficients decrease in time, with petroleum at the bottom, in qualitative agreement with Fig.~\ref{Fig:Gini}.

\section{Discussion and Conclusions}

The cost and strategies for climate change mitigation are often discussed in the economic literature using a representative agent model \cite{Rezai-2012}, where the entire population is reduced to a single ``typical'' agent.  By construction, such approaches ignore the enormous heterogeneity and inequality of living conditions of the world population.  On the other hand, there is also growing recognition that inequality by itself is a serious obstacle for further progress of human society \cite{Stiglitz-2012}.  For example, international talks on climate change for the last 20 years failed to produce any result \cite{IEA}, in part because vital economic interests of the countries at the opposite ends of the energy consumption distribution, such as the USA and India, are so different that it is hard for these countries to agree on substantive policy measures.  It may be argued that overcoming global inequality should be a goal in itself \cite{Stiglitz-2012}, which would then facilitate global agreements on climate change and other issues.  Our paper shows that global inequality in energy consumption and carbon emissions per capita has, indeed, decreased significantly in the last 30 years, presumably due to the globalization of the world economy.  However, we also show that global distribution is now close to exponential, as expected from the principle of maximal entropy.  Thus, we argue that the global economy will slow down and that global inequality will stay put at approximately the present level, because the global probability distribution of energy resources is already close to maximal entropy.  Furthermore, there are arguments in the literature for distinguishing between equality and fairness.  An equal distribution in not necessarily fair, and Ref.~\cite{Venky-2009,Venky-2010} argues that a fair distribution is the one that maximizes entropy.  A similar idea was advanced in Refs.\ \cite{Yakovenko-2000,Yakovenko-RMP,Yakovenko-2011,Banerjee-2010,Yakovenko-2013} by arguing that an exponential distribution represents a ``natural'' inequality, not in the sense that it is fair, but in the sense that it is virtually unavoidable, because it is difficult to fight against entropy maximization.

The important conclusion of our study is that, as long as fossil fuels are redistributable around the world, the broad exponential inequality in energy consumption and concomitant standards of living seems to be inevitable.  This stubborn inequality greatly complicates any global agreement on limiting carbon emissions because of the vast difference between the countries at the opposite ends of the spectrum.  One way out of this conundrum may be significant reorientation of the global economy toward renewable energy.  Not only would this lower carbon emissions, but it would also reduce inequality, assuming that solar and wind resources are distributed more uniformly around the world.  The principle of maximal entropy would not apply to locally generated and locally consumed decentralized renewable energy, because it is not easily redistributable across long distances on global scale.  Switching to renewable energy may be a way for developing countries to improve their standards of living, because these countries would not be able to increase their share of fossil fuel consumption given the exponential inequality pattern for redistributable resources.

\acknowledgements{Acknowledgments} 
This work was supported by a grant from the Institute for New Economic Thinking (INET).  We thank Branko Milanovi\'c (Lead Economist of the Development Research Group in the World Bank) for sending us Refs.~\cite{Milanovic-apart,Milanovic-2012}.

\conflictofinterests{Conflicts of Interest}
The authors declare no conflict of interest. 

\bibliographystyle{mdpi}

\begin{thebibliography}{99}

\bibitem{EIA} U.S.\ Energy Information Administration (EIA): International Energy Statistics. Available online:
\url{http://www.eia.gov/cfapps/ipdbproject/iedindex3.cfm} (accessed on 12 December 2013).  We downloaded the tables for years 1980--2010, for all countries (but not regions), for population, $\rm CO_2$ emissions, and total energy consumption.

\bibitem{trillion}  Allen, M.R.;\ Frame, D.J.;\ Huntingford, C.;\ Jones, C.D.;\ Lowe, J.A.;\ Meinshausen, M.;\ Meinshausen, N. Warming caused by cumulative carbon emissions towards the trillionth tonne. {\em Nature} {\bf 2009}, {\em 458}, 1163--1166, \url{http://dx.doi.org/10.1038/nature08019}.

\bibitem{IPCC} Intergovernmental Panel on Climate Change (IPCC).  Available online: \url{http://www.climatechange2013.org} (accessed on 12 December 2013).

\bibitem{NYTimes} Gillis, J. U.N.~climate panel endorses ceiling on global emissions. {\em The New York Times}, 27 September 2013.  Available online: \url{http://www.nytimes.com/2013/09/28/science/global-climate-change-report.html} (accessed on 12 December 2013).

\bibitem{Yakovenko-Web} Yakovenko, V.M. Econophysics Research in Victor Yakovenko's Group.  Available online: \url{http://physics.umd.edu/~yakovenk/econophysics/} (accessed on 12 December 2013).

\bibitem{Wannier} Wannier, G.H. {\em Statistical Physics}; Dover: New York, 1987.

\bibitem{Gorban} Gorban, A.N.  \textit{Maxallent}: Maximizers of all entropies and uncertainty
of uncertainty.  {\em Comput.~Math.~Appl.}\ {\bf 2013}, {\em 65}, 1438--1456, \url{http://dx.doi.org/10.1016/j.camwa.2013.01.004}.

\bibitem{Yakovenko-2000} Dr\u{a}gulescu, A.A.;\ Yakovenko, V.M.  Statistical mechanics of money. {\em Eur.~Phys.~J.~B} {\bf 2000}, {\em 17}, 723--729, \url{http://dx.doi.org/10.1007/s100510070114}.

\bibitem{Yakovenko-RMP} Yakovenko, V.M.;\ Rosser, J.B.  Colloquium: Statistical mechanics of money, wealth, and income. {\em Rev.~Mod.~Phys.}\ {\bf 2009}, {\em 81}, 1703--1725, \url{http://dx.doi.org/10.1103/RevModPhys.81.1703}.

\bibitem{Yakovenko-2011} Yakovenko, V.M.  Statistical mechanics approach to the probability distribution of money. 
In {\em New Approaches to Monetary Theory: Interdisciplinary Perspectives}; Ganssmann, H.,\ Ed.; Routledge: London, 2011; pp.\ 104--123, ISBN 978-0-415-59525-4, \url{http://arxiv.org/abs/1007.5074}.

\bibitem{Yakovenko-2001a} Dr\u{a}gulescu, A.A.;\ Yakovenko, V.M.  Evidence for the exponential distribution of income in the USA. {\em Eur.~Phys.~J.~B}, {\bf 2001}, {\em 20}, 585--589, \url{http://dx.doi.org/10.1007/PL00011112}.

\bibitem{Yakovenko-2001b} Dr\u{a}gulescu, A.A.;\ Yakovenko, V.M.  Exponential and power-law probability distributions of wealth and income in the United Kingdom and the United States. {\em Physica A}, {\bf 2001}, {\em 299}, 213-221, \url{http://dx.doi.org/10.1016/S0378-4371(01)00298-9}.

\bibitem{Yakovenko-2006} Banerjee, A.;\ Yakovenko, V.M.;\ Di Matteo, T.  A study of the personal income distribution in Australia. {\em Physica A}, {\bf 2006}, {\em 370}, 54--59, \url{http://dx.doi.org/10.1016/j.physa.2006.04.023}.

\bibitem{Derzsy-2012} Derzsy, N.;\ N\'eda Z.;\ Santos, M.A.  Income distribution patterns from a complete social security database. {\em Physica A}, {\bf 2012}, {\em 391}, 5611--5619, \url{http://dx.doi.org/10.1016/j.physa.2012.06.027}.

\bibitem{Jagielski-2013} Jagielski, M.;\ Kutner, R.  Modelling of income distribution in the European Union with the Fokker-Planck equation. {\em Physica A}, {\bf 2013}, {\em 392}, 2130--2138, \url{http://dx.doi.org/10.1016/j.physa.2013.01.028}.

\bibitem{Yakovenko-2003} Dr\u{a}gulescu, A.A.;\ Yakovenko, V.M.  Statistical mechanics of money, income, and wealth: a short survey.  In {\em Modeling of Complex Systems: Seventh Granada Lectures}; Garrido, P.L.,
Marro, J., Eds.; {\em AIP Conf.~Proc.}\ {\bf 2003}, {\em 661}, 180--183, \url{http://dx.doi.org/10.1063/1.1571309}.

\bibitem{Silva-2005} Silva, A.C.;\ Yakovenko, V.M. Temporal evolution of the `thermal' and `superthermal' income classes in the USA during 1983--2001. {\em Europhys.~Lett.}\ {\bf 2005}, {\em 69}, 304--310, \url{http://dx.doi.org/10.1209/epl/i2004-10330-3}.

\bibitem{Banerjee-2010} Banerjee, A.;\ Yakovenko, V.M.  Universal patterns of inequality. {\em New J.~Phys.}\ {\bf 2010}, {\em 12}, 075032, \url{http://dx.doi.org/10.1088/1367-2630/12/7/075032}.

\bibitem{Chakrabarti-2013} Chakrabarti, B.K.;\ Chakraborti, A.;\ Chakravarty, S.R.;\ Chatterjee, A.  {\em Econophysics of Income and Wealth Distributions};  Cambridge University Press: Cambridge, 2013, ISBN 9781107013445.

\bibitem{Milanovic-2012} Milanovi\'c, B.  Global inequality recalculated and updated: the effect of new PPP estimates on global inequality and 2005 estimates. {\em J.~Econ.~Inequal.} {\bf 2012}, {\em 10}, 1--18, \url{http://dx.doi.org/10.1007/s10888-010-9155-y}.

\bibitem{Milanovic-apart} Milanovi\'c, B. {\em Worlds Apart: Measuring International and Global Inequality};  Princeton University Press: Princeton, 2007, ISBN 9780691130514.

\bibitem{Lorenz} Sauerbrei, S.  Lorenz curves, size classification, and dimensions of bubble size distributions.  {\em Entropy} {\bf 2010}, {\em 12}, 1--13, \url{http://dx.doi.org/10.3390/e12010001}.

\bibitem{Dent} Dent Jr.,\ H.S.; Johnson, R. {\em The Great Crash Ahead}; Free Press: New York, 2011, ISBN 978-1451641554.

\bibitem{Ice-Age} Bott, U.  The coming global economic ice age? {\em The Globalist}, 12 August 2013. Available online: \url{http://www.theglobalist.com/the-coming-global-economic-ice-age/} (accessed on 12 December 2013).

\bibitem{India} Gowen, A.  In India, economic slowdown and inflation cause middle class to defer dreams.  {\em Washington Post}, 5 November 2013. Available online:  \url{http://wapo.st/16DTJ12} (accessed on 12 December 2013).

\bibitem{Yakovenko-2013} Yakovenko, V.M.  Applications of statistical mechanics to economics: Entropic origin of the probability distributions of money, income, and energy consumption.  In {\em Social Fairness and Economics: Economic essays in the spirit of Duncan Foley}; Taylor, L., Rezai, A., Michl, T., Eds.; Routledge: London, 2013, pp.\ 53--82, ISBN 978-0-415-53819-0,
\url{http://arxiv.org/abs/1204.6483}.

\bibitem{Friedman-flat} Friedman, T.L. {\em The World Is Flat}; Farrar, Straus and Giroux: New York, 2005, ISBN 0374292884.

\bibitem{Pareto-20-80} Pareto, V. {\em 1906 Manuale di Economia Politica}; EGEA -- Universit\`a Bocconi Editore: Milan, 2006, ISBN 88-8350-084-9.

\bibitem{Rezai-2012} Rezai, A.;\ Foley, D.K.;\ Taylor, L.  Global warming and economic externalities. {\em Econ.~Theory} {\bf 2012}, {\em 49}, 329--351, \url{http://dx.doi.org/10.1007/s00199-010-0592-4}.

\bibitem{Stiglitz-2012} Stiglitz, J.E., {\em The Price of Inequality: How Today's Divided Society Endangers our Future}; Norton: New York, 2012, ISBN 978-0393088694.

\bibitem{IEA} Bakewell, S. Energy as dirty as 20 years ago on slow climate effort, IEA says. {\em Bloomberg News}, 17 April 2013. Available online: \url{http://www.bloomberg.com/news/2013-04-17/energy-as-dirty-as-20-years-ago-on-slow-climate-effort-iea-says.html} (accessed on 12 December 2013).

\bibitem{Venky-2009} Venkatasubramanian, V.  What is fair pay for executives? An information theoretic analysis of wage distributions.  {\em Entropy} {\bf 2009}, {\em 11}, 766--781; \url{http://dx.doi.org/10.3390/e11040766}. 

\bibitem{Venky-2010} Venkatasubramanian, V.  Fairness is an emergent self-organized property of the free market for labor.  {\em Entropy} {\bf 2010}, {\em 12}, 1514--1531, \url{http://dx.doi.org/10.3390/e12061514}.

\end{thebibliography}
\makeatletter
\renewcommand\@biblabel[1]{#1.}
\makeatother

\end{document}